\documentstyle[emulateapj]{article}

\def\h0{$H_{0}=50$~km~sec$^{-1}$~Mpc$^{-1}$}
\def\ergssec   {~ergs~sec$^{-1}$}
\def\msun     {$M_{\odot}$}
\newcommand{\keV}{ke\kern-0.05em V}

\begin{document}

\lefthead{VIKHLININ, FORMAN, AND JONES}
\righthead{ANOTHER COLLISION FOR  THE COMA CLUSTER}

\slugcomment{submitted to {\em The Astrophysical Journal Letters}}

\title{Another Collision for the Coma Cluster}

\author{A.~Vikhlinin, W.~Forman, and C.~Jones}
\affil	{Harvard-Smithsonian Center for Astrophysics, 60 Garden street,
 Cambridge, MA 02138, USA.\\ e-mail: avikhlinin@cfa.harvard.edu,
wforman@cfa.harvard.edu, cjones@cfa.harvard.edu}

\begin{abstract}

We describe a wavelet transform analysis of the ROSAT PSPC images of the
Coma cluster. On small scales, $\le 1^\prime$, the wavelet analysis shows
substructure dominated by two extended sources surrounding the two brightest
cluster galaxies NGC 4874 and NGC 4889.  On slightly larger scales, $\sim
2'$, the wavelet analysis reveals a filament of X-ray emission originating
near the cluster center, curving to the south and east for $\sim 25^\prime$
in the direction of the galaxy NGC 4911, and ending near the galaxy NGC
4921.  These results extend earlier ROSAT observations and further indicate
the complex nature of the cluster core.  We consider two possible
explanations for the production of the filamentary feature as arising from
interactions of the main cluster with a merging group. The feature could
arise from either ram pressure stripped gas or a dark matter perturbation of
tidally stripped material.

\end{abstract}                                          


\section{Introduction}

Clusters of galaxies have relatively long dynamical time-scales, comparable
to the age of universe, and hence can provide insights into the conditions
and properties of the universe from which these massive systems arose. The
study of substructure in clusters has been applied to deriving cosmological
parameters (Richstone, Loeb \& Turner 1992; Kaufman \& White 1993; Lacey \&
Cole 1993; Mohr et al.\ 1995).  In principle, statistical studies of
substructure can provide strong constraints on the density of the Universe,
although uncertainties remain including the time-scale for substructure to
be erased and the nature of the initial mass fluctuation spectrum (Kaufman
\& White 1993).

The Coma cluster, as the second X-ray brightest cluster and one of the
nearest rich clusters, can provide insights into the reliability of
estimates of $\Omega$ from substructure studies. Long-considered as the
typical dynamically relaxed system, X-ray studies have shown its remarkable
complexity (Briel et al.\ 1992; White, Briel \& Henry 1993; Vikhlinin,
Forman \& Jones 1994).  Optical indications of substructure come from the
work of Fitchett \& Webster (1987) who demonstrated the presence of
substructure and Mellier et al.\ (1988) who were able to derive the mass to
light ratios for the core substructures surrounding the two dominant core
galaxies. Most recently, Colless and Dunn (1995) used a sample of 552 galaxy
redshifts to further clarify the merger history of the Coma cluster.
Biviano et al.\ (1996) combined ROSAT and extensive optical redshift and
photometric samples to identify the main body of the cluster and suggested
that the main cluster body was undergoing rotation. They also concluded,
based on the difference in velocity between the NGC 4889 and NGC 4874 groups
and the cluster mean, that these groups only recently arrived within the
cluster core.

We have revisited the extensive ROSAT PSPC archival data for the Coma
cluster using a new technique, particularly well-suited to detecting small
scale structures embedded within larger scale features.  This technique,
``wavelet transform decomposition'', allows simultaneous study of the image
at all spatial (angular) scales and allows identification of the scale
appropriate to the detected features.  In this {\em Letter}, we study the
small scale structure of the Coma image with this new wavelet transform
technique.  We detect a long, narrow X-ray feature and discuss possible
explanations as either gas which was ram pressure stripped from a small
galaxy group during a recent merger/collision with the cluster or a dark
matter perturbation produced by tidally stripped material during a group
merger.

\section{OBSERVATIONS AND ANALYSIS}

We analyzed two ROSAT PSPC pointings of the Coma cluster core (the overall
X-ray morphology of the Coma cluster seen from pointed ROSAT observations is
discussed in White et al.\ 1993).  Images were reduced using the
prescriptions and code of Snowden et al.\ (1994).  This includes the
elimination of high background intervals, which are primarily due to intense
scattered solar X-rays, subtraction of the particle and other non-cosmic
background components\footnote{The so-called ``long term enhancement'' (see
explanations in Snowden et al.) and the residual solar scattered X-rays},
and creation of the exposure maps appropriate for the given energy band
which are used for flat-fielding and merging the images obtained from
different pointings. The cleaning resulted in 18 and 13 ksec exposures in
individual pointings. The merged PSPC image (top-left panel in Fig~1) shows
an azimuthally asymmetric cluster (Briel at al.\ 1992, White et al.\ 1993)
whose core structure we investigate in detail in this {\em Letter}.

\subsection{Wavelet Decomposition}

The technique we applied to the ROSAT PSPC ``cleaned'' images employs a set
of {\em \`a trous}\/ wavelet transforms, closely following the approach of
Starck et al.\ (1995 and references therein) with some important changes in
the iteration scheme. An application of the multi-resolution wavelet-based
image filtering for the X-ray images also can be found in Slezak et al.\
(1994).

The {\em \`a trous}\/ wavelet transform of scale $a$ can be (very roughly)
characterized as the convolution of an image with the function equal to the
difference between two Gaussians, the first positive one with $\sigma=a$,
and the second negative one with $\sigma=2a$. The basic idea of the
wavelet-based image filtering is to perform the set of wavelet transforms
with scales $a, 2a, 4a ...$, retain only significant wavelet coefficients at
each scale (thus filtering out the noise), and perform the image
reconstruction from the wavelet planes. For the {\em \`a trous}\/ wavelet
transform, the reconstruction is performed by simple summation of all
wavelet planes.  At each scale we retain the wavelet coefficients which
exceed some critical amplitude which corresponds to the given number of
standard deviations ($3\sigma$ in our application). Since the wavelet
transform is linear, the noise level in each wavelet plane is proportional
to the noise level in the original image. Therefore, we can derive the
critical amplitudes for each scale using the following approach. We
first apply the wavelet transform to a simulated image containing 
Gaussian noise with dispersion 1 and determine the $rms$ variations in the
wavelet-convolved images. We then multiply these values by the local noise
in the actual image, which is estimated using the local background 
and exposure.
Thus, the faintest detectable structures have uniform amplitude in terms of
significance, as a function of both position and scale.

\begin{figure*}[htb]
\centerline{\hbox{ \epsfysize=3.0in \epsffile{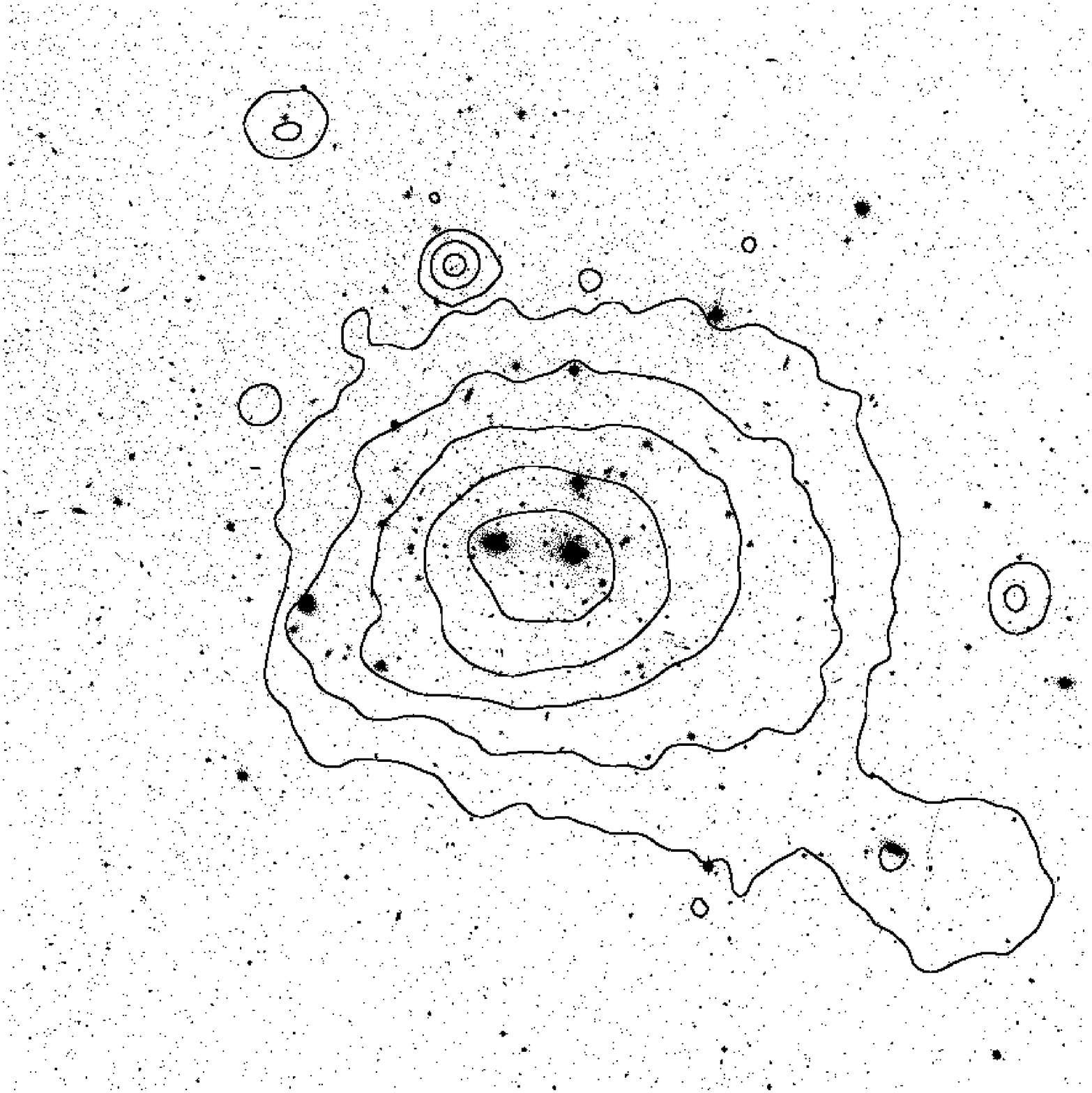}
       \epsfysize=3.0in \epsffile{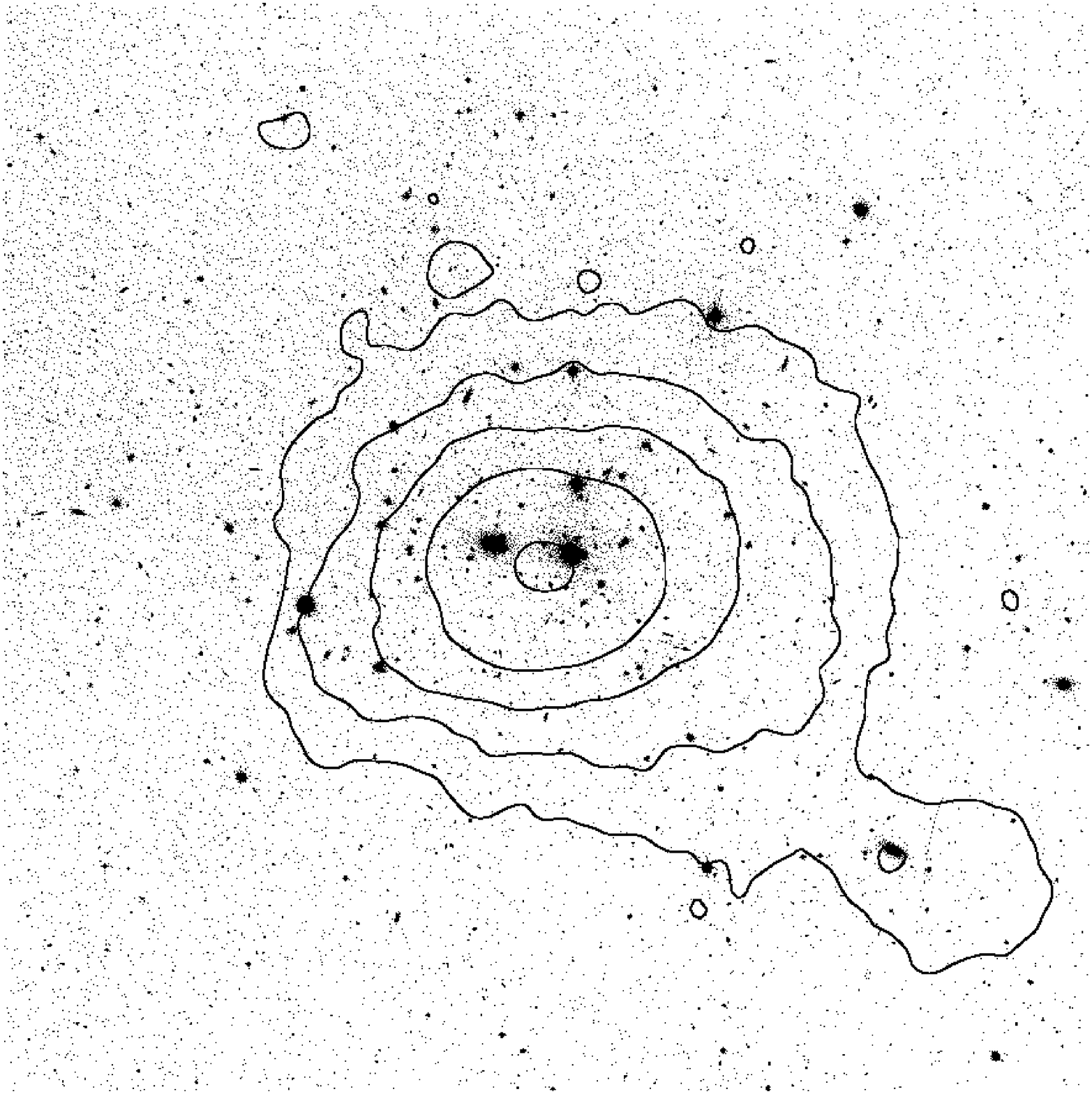}}}
\centerline{\hbox{\epsfysize=3.0in \epsffile{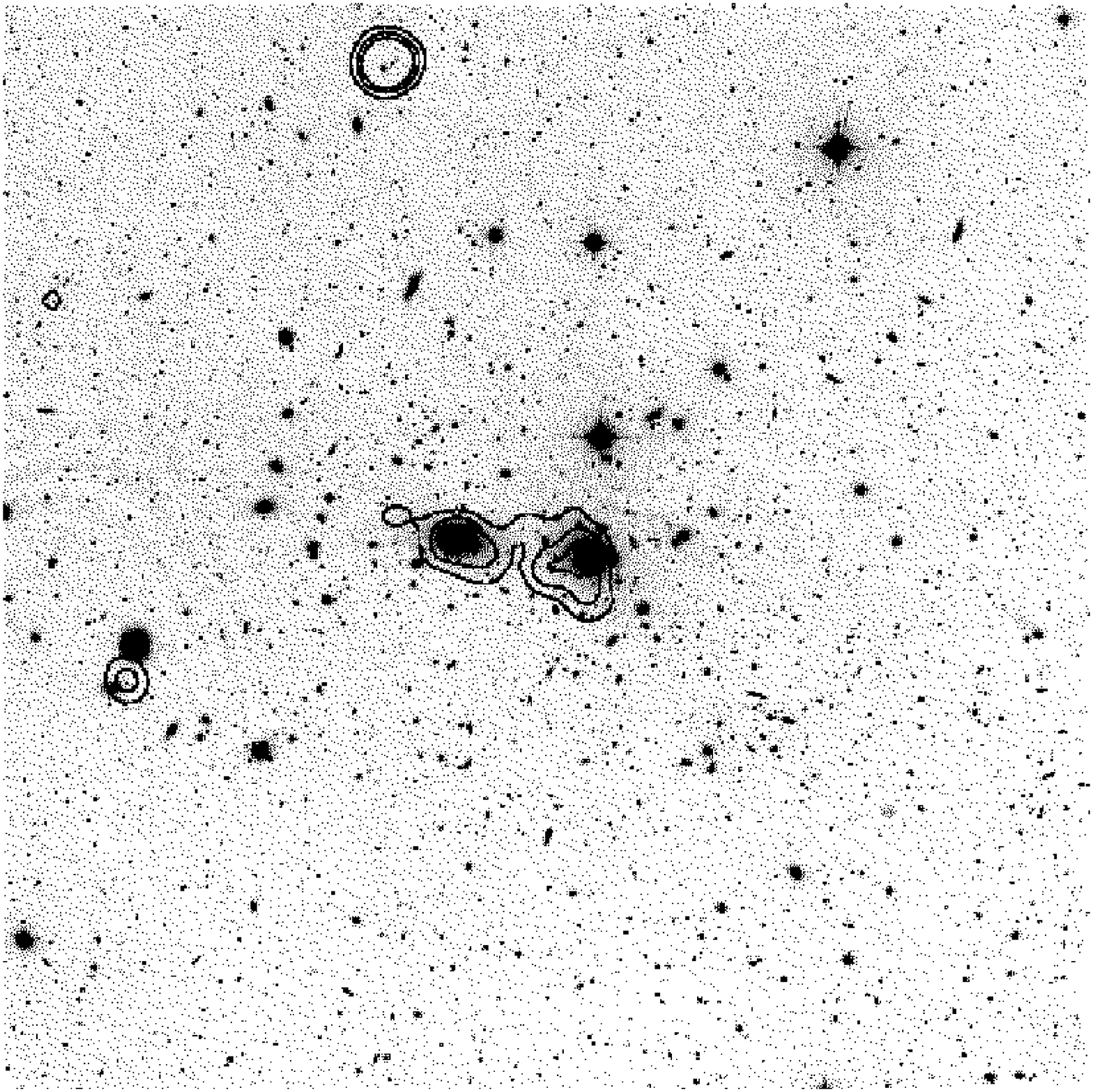} 
      \epsfysize=3.0in \epsffile{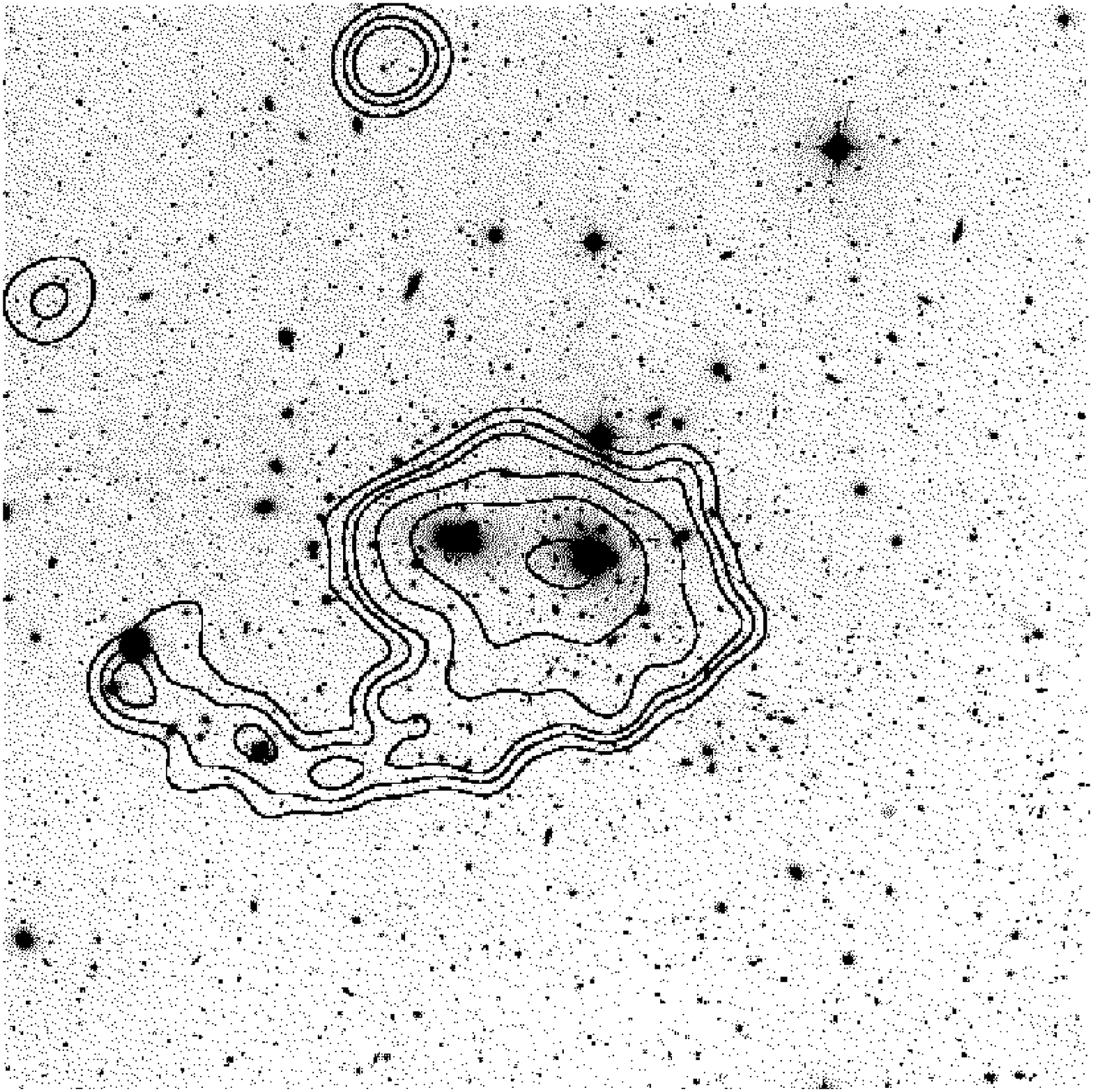}}}

\vspace{1mm}
\small\parindent=3.5mm
{\sc Fig.}~1.---({\em top-left}) The X-ray image of the Coma cluster in the
energy band 0.5--2 \keV, smoothed with the Gaussian $\sigma=1'$, overlayed
on the optical image. ({\em bottom-left}) Structure detected in the Coma
cluster core on the wavelet scale $60''$ The most prominent features are
extended sources around NGC 4889 and NGC 4874, as well as a few point
sources.  ({\em bottom-right}) Structure detected on the $120''$ scale. The
tail of enhanced emission originates near NGC 4874, goes to the South and
East towards NGC 4911 and ends near NGC 4921. The formal statistical
significance along the axis of the tail varies from $4.5\sigma$ to
$6\sigma$, and the surface brightness is $\sim 10$\% of the underlying
cluster emission.  ({\em top-right}) Small scale structures shown in bottom
panels were subtracted from the Coma cluster image, and the result was
smoothed with the Gaussian $\sigma=1'$. After the subtraction of the
small-scale structure the cluster core appears much more regular. The
complex morphology (elongation, and the rotation of the elongation axis)
therefore reflects the ``hidden'' small-scale structure, not the state of
the main cluster body.
\end{figure*}

To assure the most complete separation of image features in terms of their
scale, we apply the following iterative procedure. We start from the
smallest scale, $a=30''$, which roughly corresponds to the FWHM of the ROSAT
PSPC PSF. We calculate the wavelet transform on this first scale only, find
significant features, and subtract them from the original image (recall that
the back-transformation for the {\em \`a trous}\/ wavelet transform is
simply the summation of the wavelet planes). The next iteration is applied
to the residual and this process is repeated until no new significant
features are found at this scale. We then start the process with the scale
$2a$ and so on. The subtraction of the features detected at smaller scales
before applying the wavelet transform at larger scales results in a
sufficient enhancement of the basic property of the wavelet transform
analysis --- its ability to separate image features in terms of their
characteristic scale.  In particular, we are able to isolate even extremely
bright point sources primarily at the two smaller scales (i.e.\ $\leq
60''$), whereas without subtraction, they appear at all scales and often
saturate nearby and relatively faint extended structures.

The process described above provides a separation of the original image into
a set of images containing only features of scale $a$, $2a$, ...
significant at least at the $3\sigma$ level, which can be either summed
together to produce an adaptively smoothed image or viewed individually (or
in any desired combination) to examine the significant structures at desired
scales.

\subsection{Small Scale Structure of the Coma Cluster}

Contour maps of the images generated for small scales are shown in the
bottom panels of Fig~1 superposed on an optical image.  On the smallest
scales, $60''$ ({\em bottom-left}), the contours show the extended sources
surrounding the two brightest cluster galaxies NGC 4874 and NGC 4889 first
detected in X-rays by White et al.\ (1993; see also Vikhlinin et al.\ 1994).
On larger scales, $120\arcsec$ ({\em bottom-right}), the wavelet analysis
reveals a filamentary X-ray emission feature originating near the cluster
center, curving to the south and east for $\sim25'$ towards the galaxy NGC
4911, and ending near the galaxy NGC 4921. The surface brightness
enhancement near NGC 4911 was first noted by White et al.\ (1993).

Within the outer contour of the filament, the surface brightness ranges from
5--10\% of the underlying diffuse emission from Coma itself.  The formal
significance along the axis of the filament varies from $4.5\sigma$ to
$6\sigma$. Features with this significance never appear in simulated Coma
cluster images. The counts associated with the emission filament (within the
outer contour shown in bottom-right panel of Fig~1) correspond to a count
rate of $0.12$ cts s$^{-1}$ in the energy band 0.5--2 \keV.  The
corresponding luminosity, for \h0 and a sufficiently hard spectrum (see
below) is $3\times10^{42}$ \ergssec.  At the Coma distance ($z=0.0235$), the
tail is approximately $1$ Mpc (25\arcmin) in length with a diameter of
$160$ kpc (4\arcmin).


It is interesting to constrain the spectral properties of the filamentary
structure. It is faint compared to the underlying cluster emission, which
therefore cannot be neglected and must be carefully subtracted. We use the
following technique for the extraction of the filament's spectrum. We
perform the wavelet decomposition of the broad band image as described above
and record all the transformations which have been applied. The identical
transformations are performed to the images in individual energy bands
defined by the energies 0.2--0.4--0.5--0.7--0.9--1.3--2 \keV. Since the
wavelet transform is linear, the relative fluxes in different energy bands
are conserved and can be used to fit the spectrum.  Unfortunately, there is
a serious problem with the PSPC spectral calibration for these particular
observations, which results in anomalously low temperatures for the total
cluster emission.\footnote{To our knowledge, a similar effect is found for
other clusters observed around July 1991: A3558 (Markevitch \& Vikhlinin
1997) and A1795 (L.\ David, private communication), have anomalously low
best fit temperatures. This suggests that a likely explanation is a detector
gain variation around July 1991, which is not accounted for by the present
versions of the PSPC calibration.} We therefore use the ratio of the
filament brightness to the underlying cluster brightness as a function of
energy to constrain the filament spectrum. This ratio is constant within the
errors, which suggests that the filament has the same spectrum as the Coma
cluster. Quantitatively, a 95\% lower limit to the temperature in the
filament is $4$ \keV, assuming a cluster core temperature of $8$ \keV\
(Hughes, Gorenstein \& Fabricant 1988).

\section{DISCUSSION}

The wavelet analysis of the ROSAT PSPC X-ray images of the Coma cluster
demonstrates the existence of a new feature possibly associated with
dynamical activity in the cluster core. We can estimate the physical
parameters associated with this filament of emission assuming it arises from
hot gas.  Let us assume that we are viewing the filament in the plane of the
sky and it is a uniform cylinder of radius $100$ kpc and length $1$ Mpc.
The observed 10\% surface brightness enhancement implies that the density of
the filament must exceed the cluster gas density at least by about 5\%, if
the gas is not particularly cool ($T>$ a few \keV). The relative brightness
of the filament does not change much along its axis, therefore the gas
density enhancement in the filament is an almost constant small perturbation
and one can use the main cluster density profile (Briel et al.\ 1992) to
estimate the filament's gas mass. We derive a total average gas density in
the volume defined by the filament of approximately $10^{-3}$ cm$^{-3}$
with a range (along the filament's axis) from about $2.5\times10^{-4}$ to
$2.5\times10^{-3}$ cm$^{-3}$. This density translates into a gas mass of
$5\times10^{11}$ \msun. We consider two possible scenarios for the origin
of this filamentary feature.

First, let us suppose that this feature arises from gas that has been ram
pressure stripped from a merging group. The observed gas mass is well within
the reasonable range for the gas mass in the core of a group (see for
example, Fig~4 of David et al.\ 1995).  More importantly, however, we must
consider whether such a thin structure could survive in the cluster
environment.  Following Cowie \& McKee (1977), we have computed the
evaporation rate for a spherical cloud embedded in the hot ICM of the Coma
cluster. In the Coma cluster core, the classical thermal conduction should
provide a reasonable estimate for the evaporation time, which is given as
\begin{equation}
t_{\rm evap} = 3.3\times10^{20} n_c R^2_{\rm pc} T_f^{-5/2} (\ln \Lambda / 30)
\, {\rm yr},
\end{equation}
where $n_c$ is the embedded cloud density, $R_{\rm pc}$ is its radius in pc,
$T_f$ is the temperature of the surrounding, hot medium, and $\Lambda$ is
the Coulomb logarithm. For a cloud of radius $10^2$ kpc, density
$2.5\times10^{-3}$ cm$^{-3}$, embedded in a hot medium of temperature
$8$ \keV, the evaporation time is $t_{\rm evap}\sim0.1$ Gyr.  In $0.1$ Gyr,
a galaxy or group traveling at the cluster velocity dispersion could
traverse $0.2$ Mpc.  The evaporation time at the end of the filament away
from the cluster center where its density is lower is significantly shorter.
In principle, cylindrical geometry makes the evaporation time longer by a
factor of a few.  Also, any chaotic magnetic field could significantly
reduce the thermal conduction efficiency and increase the evaporation
time-scale. Finally, we do not necessarily need the evaporation time-scale
to be smaller or comparable to the traverse time: if the amount of gas
deposited into the tail significantly exceeded $5\times10^{11}$ \msun, it
may be mostly evaporated at present, with only the remnants of the group gas
providing the observed surface brightness enhancement.

An alternative is that the observed filamentary feature arises from cluster
gas compressed in the potential well of the dark matter perturbation formed
by a tidally disrupted infalling galaxy group. Merritt (1984) has shown that
galaxy halos are tidally disrupted by the cluster gravitational field.
Specifically, a gravitationally bound system entering a cluster is tidally
limited at a radius $r_t$ given by:
\begin{equation}
r_t/R_c = F(r/R_c) \sigma_g/\sigma_c
\end{equation}
where $R_c$ is the cluster core radius, $\sigma_g$ is the group velocity
dispersion, $\sigma_c$ is the cluster velocity dispersion, $r$ is the
distance from the cluster center, and $F(r/R_c)$ is a function which has a
minimum value of 1/2 at $r=R_c$ (see Fig~1 in Merritt 1984).  The function
$F(r/R_c)$ decreases nearly linearly with decreasing radius and hence an
infalling group will be tidally disrupted as it approaches $r=R_c$.  The
minimum size depends in detail on the parameters of the group and cluster,
but for a typical group with $\sigma_g=300$ km sec$^{-1}$ in the Coma
cluster with $\sigma_c=1142$ km sec$^{-1}$ (Biviano et al.\ 1996) and
$R_c=420$ kpc (Briel et al.\ 1992), the maximum tidal radius for an
infalling group is $r_t\sim50$ kpc. Thus, the group will be progressively
disrupted as it approaches the core. A trail of tidally stripped debris
behind the group provides the gravitational potential perturbation which
enhances the surface brightness. For isothermal gas in hydrostatic
equilibrium, the gas density perturbation is:
\begin{equation}
(\rho_0 + \delta \rho)/\rho_0 = {\rm exp} ( -\mu m_p \Delta\varphi/kT).
\end{equation}
where $\rho_0$ is the initial gas density, and $kT$ is the gas temperature.
The observed surface brightness enhancement corresponds to a gas density
enhancement of about 5\%, hence the change in the gravitational potential,
$\Delta\varphi$, can be written as $1.05 = {\rm exp} (-\mu
m_p\Delta\varphi/kT)$ or $\Delta\varphi = 0.05 kT/\mu m_p$.

The gravitational potential on, and near, the surface of a long cylinder is:
\begin{equation}
\psi = - G \rho_l \ln(L/R)
\end{equation}
where $\rho_l$ is the linear mass density, $L$ is the length of the
cylinder, and $R$ is its radius.  If we equate the potential on the surface
of the cylinder to the change in the gravitational potential required to
produce the observed surface brightness enhancement, then we can solve for
the linear mass density in the cylinder. This provides an expression for the
added mass in the cylinder:
\begin{equation}
\rho_l L = \frac{\Delta\varphi L}{G \ln(L/R)} = 
\frac{0.05 kT L}{\mu m_p G \ln (L/R)}
\end{equation}
For the observed properties of the filament, and assuming the gas
temperature is that of the cluster, we find a mass of $M=6\times10^{12}$
\msun. At most, the mass could be 50\% smaller if the temperature
of the gas in the filament is at its measured lower limit. The surface
brightness enhancement also may be underestimated by a factor of a few,
since the filament may not be located in the densest part of the projected
gas distribution in which case the required density contrast is larger.
Also, the 10\% enhancement in the surface brightness corresponds to more
than 10\% enhancement in the volume emissivity, because the filament is
narrow and projected onto the main cluster emission. Thus, the measured mass
in the perturbation is approximately $6\times10^{12}$ \msun\ and is uncertain
by a factor of about 2. This mass is well within the range of that in the
central region of a group (e.g.\ David et al.\ 1995).

We have discussed the filament in some detail, but the remaining emission is
also of interest. The top-right panel in Fig~1 compares the ``raw'' smoothed
image of Coma ({\em top-left}) with the smoothed image after subtraction of
the detected small scale structure ({\em bottom} panels).  The removal of
the two central X-ray peaks around the galaxies NGC 4889 and NGC 4874, as
well as the filament, leaves a smooth and regular-looking X-ray surface
brightness distribution.  A relatively small surface brightness perturbation
can produce marked irregularity in the surface brightness distribution.
Therefore, the small scale structure can significantly influence the overall
appearance of the surface brightness distribution of the cluster.  Numerical
simulations must retain sufficient resolution, if they are to fully model
shapes of the X-ray cluster images.

\section{CONCLUSIONS}

We report the detection of a long, linear filamentary feature extending over
approximately $1$ Mpc from the Coma cluster center toward NGC 4911.  We
explore two possibilities for the origin of this filament.  First, we show
that the filament could arise from ram pressure stripped gas from a group
passing through Coma, especially if magnetic fields significantly reduce the
conduction efficiency and lengthen the evaporation time.  Second, we note
that tidally stripped dark matter from a merging group could produce a
perturbation that would give rise to the observed structure.

Increases in sensitivity and new analysis techniques have continued to
demonstrate the complexity of the cluster environment.  The Coma cluster,
long thought to be the prototype for a relaxed cluster, is in fact a
remarkably complex system with structure on many different scales seen both
optically and in X-rays.

\acknowledgments

We thank L. David for helpful discussions and acknowledge support from
NAS8-39073.

%
%
%

\end{document}